\documentclass[twocolumn,prb,superscriptaddress,aps,10pt,bibnotes,amsfonts,amsmath,amssymb,floatfix]{revtex4-2} 

\usepackage{graphicx}
\usepackage{ifthen}
\usepackage{xcolor}
\usepackage{bm}
\usepackage{braket}
\usepackage{multirow}
\usepackage{hyperref}
\usepackage{soul}
\usepackage{svg}
\usepackage{ulem}

\hypersetup{
 pdfnewwindow=true, colorlinks=true,
 linkcolor=blue, anchorcolor=blue,
 citecolor=blue, filecolor=blue,
 menucolor=blue, urlcolor=blue}

\def\bk{\mathbf{k}}

\def\bq{\mathbf{q}}

\def\ve{\varepsilon}
\def\ef{\ve_{\rm F}}

\begin{document}

\title{Comparative study of phonon-limited carrier transport in the Weyl semimetal TaAs family}

\author{Shashi B. Mishra}
\affiliation{Department of Physics, Applied Physics and Astronomy, Binghamton University-SUNY, Binghamton, New York 13902, USA}
\thanks{These authors contributed equally.}
\author{Zhe Liu}
\affiliation{Department of Physics, Applied Physics and Astronomy, Binghamton University-SUNY, Binghamton, New York 13902, USA}
\thanks{These authors contributed equally.}
\author{Sabyasachi Tiwari}
\affiliation{Oden Institute for Computational Engineering and Sciences, The University of Texas at Austin, Austin, Texas 78712, USA}
\affiliation{Department of Physics, The University of Texas at Austin, Austin, Texas 78712, USA}
\author{Feliciano Giustino}
\affiliation{Oden Institute for Computational Engineering and Sciences, The University of Texas at Austin, Austin, Texas 78712, USA}
\affiliation{Department of Physics, The University of Texas at Austin, Austin, Texas 78712, USA}
\author{Elena R. Margine}
\email{rmargine@binghamton.edu}
\affiliation{Department of Physics, Applied Physics and Astronomy, Binghamton University-SUNY, Binghamton, New York 13902, USA}

\begin{abstract}

We present a systematic first-principles study of phonon-limited transport in the TaAs family of Weyl semimetals using the \textit{ab initio} Boltzmann transport equation. The calculated electrical conductivities show excellent agreement with experimental data for high-quality samples, confirming that transport in these systems is predominantly limited by phonon scattering. 
Among the four compounds, NbP achieves the highest conductivity, governed primarily by its large Fermi velocities that offset its stronger scattering rates. 
In contrast, TaAs displays the lowest conductivity, linked to reduced carrier pockets and limited carrier velocities. Additionally, NbP conductivity remains largely unaffected by small hole or electron doping, whereas TaAs exhibits pronounced electron-hole asymmetry. NbAs and TaP show intermediate behavior, reflecting their Fermi surface topologies and scattering phase space. These findings provide microscopic insight into the transport mechanisms of the TaAs family and emphasize the critical role of phonons, doping, and carrier dynamics in shaping their electronic response.
\end{abstract}

\maketitle

\section{\label{sec:intro}Introduction}

The TaAs family of monopnictides was the first class of materials in which Weyl fermions were both theoretically predicted~\cite{Huang2015weyl,Weng2015} and experimentally observed~\cite{Lu2015,Xu2015,Lv2015prx,Yang2015}. 
These compounds crystallize in a body-centered tetragonal, noncentrosymmetric structure (space group $I4_1md$, No.~109), consisting of interpenetrating Ta/Nb and As/P sublattices that break inversion symmetry while preserving time-reversal symmetry. 
In the absence of inversion symmetry, spin–orbit coupling (SOC) lifts spin degeneracy away from time-reversal invariant momenta (TRIM), and the crystal symmetries constrain the number and crossing locations of nondegenerate bands near the Fermi level, producing pairs of Weyl nodes with linear dispersions. These nodes define topologically protected bulk states and give rise to surface features known as Fermi arcs. The associated band topology leads to unusual responses, 
such as the chiral anomaly~\cite{Huang2015weyl,Ma2017}, giant magnetoresistance~\cite{Ghimire2015,Arnold2016}, and ultrahigh carrier mobility~\cite{Shekhar2015, Huang2015,Zhang2017,Xiang2017}, making these materials a versatile platform for exploring topological quantum phenomena and for developing next-generation electronic and spintronic devices~\cite{Armitage2018,Nagaosa2020,Hasan2021,Rocchino2024,Zhong2025}.
While the topological aspects of the TaAs family are well established~\cite{Sun2015,Lee2015,Sun2016,Wang2016,Grassano2018, Chen2019,Zeng2021}, their charge transport properties remain a subject of active research~\cite{Liu2025,Allemand2025,Desai2025}. Transport measurements consistently reveal extremely large mobilities, sometimes exceeding $10^6$~cm$^2$V$^{-1}$s$^{-1}$ ~\cite{Huang2015,Shekhar2015,Zhang2015,Wang2016,Zhang2017,Xiang2017,Balduini2024}, but with noticeable variation across the four compounds. For example, NbP often exhibits higher mobilities than TaAs, suggesting that the strength of electron-phonon (e-ph) scattering depends sensitively on both the transition metal and the pnictogen atom. Prior studies have probed e-ph interactions through analysis of phonon linewidths, phonon transport, and optical conductivity~\cite{Coulter2019,Garcia2020,Han2023,Bonilla2024}. More recently, our work~\cite{Liu2025} and that of Allemand \textit{et al.}~\cite{Allemand2025} demonstrated that e–ph scattering is the dominant mechanism limiting electrical conductivity in TaAs, with hole doping yielding higher conductivity than electron doping. Several studies have investigated the structural, electronic, and vibrational properties across the entire TaAs family, revealing trends in Fermi-surface topology, Weyl-node positions, and Raman spectra when moving from Ta to Nb and from As to P~\cite{Lee2015,Buckeridge2016,Liu2016,Liu2017,Wang2016,Grassano2018,Chen2019,Naher2021}. However, no work has yet provided a systematic analysis of electrical conductivity across all four compounds that accounts for the observed trends or explains why NbP exhibits the highest mobility within the series. 
Here, we perform a comparative first-principles study of the temperature-dependent conductivity of TaAs, TaP, NbAs, and NbP. Our analysis indicates that e–ph scattering rates and carrier velocities are the primary factors controlling the conductivity. Their interplay results in NbP exhibiting both higher scattering rates and an enhanced average carrier velocity. Notably, in NbP, the conductivity is not affected by either electron or hole doping, whereas in the other compounds, hole carriers consistently display the highest conductivity. These findings provide detailed insight into the mechanisms governing transport in the TaAs family of Weyl semimetals and help rationalize experimental observations.
\begin{figure}[!hbt]
    \centering
    \includegraphics[width=\linewidth]{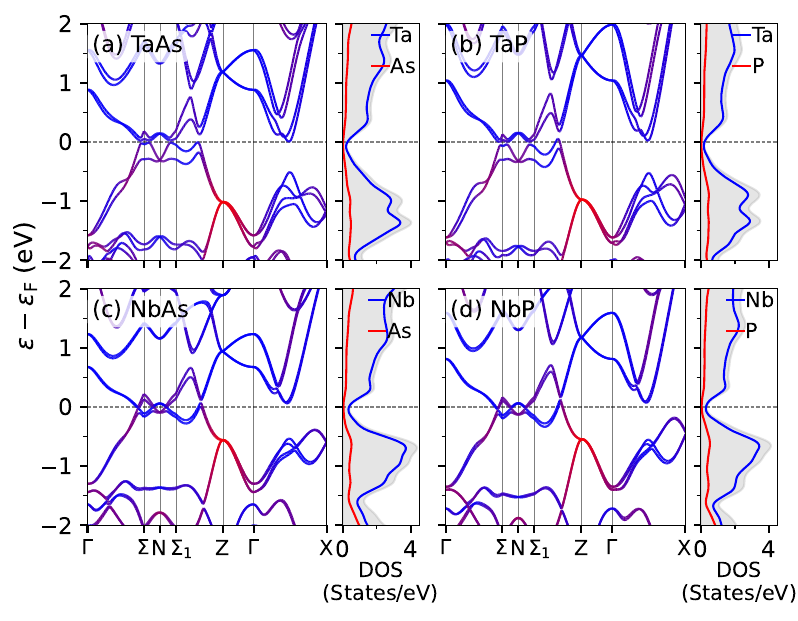}
    \caption{Electronic band structures with orbital-resolved character and corresponding total and partial density of states (DOS) for (a) TaAs, (b) TaP, (c) NbAs, and (d) NbP. Contributions from Ta/Nb and As/P orbitals are shown in blue and red colors, respectively, while the total DOS is represented by the gray shaded area.} 
    \label{fig:band}
\end{figure}

\section{Computational Details}\label{sec:methods}

The density-functional calculations are carried out using \textsc{Quantum ESPRESSO} (QE) package~\cite{Giannozzi2017}, utilizing optimized norm-conserving Vanderbilt pseudopotentials (ONCVPSP)~\cite{Hamann2013} from the Pseudo Dojo library~\cite{Vansetten2018} generated with the fully relativistic Perdew-Burke-Ernzerhof (PBE) parametrization~\cite{Perdew1996}. 
We use a plane-wave cutoff of 80~Ry, a Methfessel-Paxton smearing~\cite{Methfessel1989} value of 0.01~Ry, and a $\Gamma$-centered $12\times 12 \times 12$ $\bk$-grid. The lattice parameters and atomic positions are relaxed until the total energy is converged within $10^{-6}$~Ry and the maximum force on each atom is less than $10^{-4}$~Ry/\AA. Spin-orbit coupling is included in the electronic structure calculations. The dynamical matrices and the linear variation of the self-consistent potential due to atomic displacements are calculated within density-functional perturbation theory~\cite{Baroni2001} on a $4\times4\times4$ $\mathbf{q}$-grid. 
To investigate electron-phonon interactions and transport properties, we use the EPW code~\cite{Giustino2007,Ponce2016,Lee2023}. The electronic wavefunctions required for the Wannier interpolation~\cite{Marzari2012, Pizzi2020} are obtained on a uniform $\Gamma$-centered $8\times8\times8$ $\bk$-grid. We use 32 atom-centered orbitals to describe the electronic structure using five $d$ and three $p$ orbitals on each Ta/Nb and As/P atom as initial guesses for the Wannier functions, respectively. The iterative Boltzmann transport equations (IBTE) are solved on fine uniform $140^3$ $\bk$- and $70^3$ $\bq$-point grids, with an energy window of $\pm 0.2$~eV around the Fermi level. The Dirac $\delta$ functions appearing in the expressions for the IBTE and scattering rate are approximated by Gaussians with a broadening of 2~meV.  This set of parameters was found to give converged results in our previous study on TaAs~\cite{Liu2025}. To calculate the mode-resolved scattering rates, we used $120^3$ $\bk$- and $60^3$ $\bq$-grids, which helped mitigate computational bottlenecks.
\begin{figure}[!hbt]
    \centering
    \includegraphics[width=\linewidth]{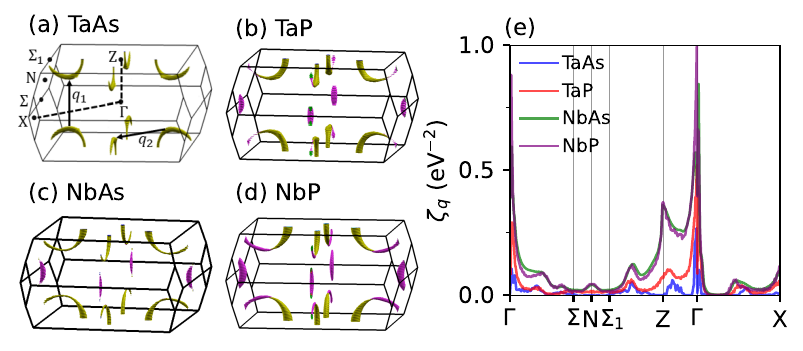}
    \caption{Fermi surfaces within a 10~meV energy window around the Fermi level for (a) TaAs, (b) TaP, (c) NbAs, and (d) NbP, generated using XCrySDen~\cite{Kokalj1999}. Yellow and magenta regions indicate hole and electron pockets. The black solid arrows in panel (a) indicate possible inter-pocket scatterings with $\bq_1$ and $\bq_2$ wave vectors along the $\Gamma$-$Z$ and $\Gamma$-$X$ directions (dashed lines). (e) Nesting functions along the same high-symmetry path for all four compounds.} 
    \label{fig:FS}
\end{figure}

\section{Results and Discussion}\label{sec:results}

The TaAs family shares the same tetragonal crystal symmetry, with structural differences arising primarily from the distinct atomic radii of Ta, Nb, As, and P. These variations lead to systematic shifts in the lattice parameters, as summarized in Table~\ref{tab:lattice_weyl}. The calculated values are in good agreement with reported lattice parameters extracted from experiments~\cite{Boller1963,Murray1976,Willerstrom1984,Xu2015}. Substituting Ta with Nb produces only a slight increase in the lattice constants, while replacing As with P results in a pronounced contraction of both $a$ and $c$. Such structural changes modify orbital hybridization and bandwidth, shifting the relative positions of Ta/Nb $d$ states and As/P $p$ states, as evident in the band dispersions of Fig.~\ref{fig:band}. In particular, consistent with prior studies~\cite{Lee2015,Grassano2018}, the Nb 4$d$ states lie closer to the Fermi level ($\ef$) than the Ta 5$d$ states along the $\Sigma$-$N$-$\Sigma_1$ path, reflecting the stronger localization of the Nb 4$d$ orbitals. Furthermore, although the $p$ orbitals contribute predominantly at energies away from $\ef$, there is significant $p$-$d$ hybridization in the vicinity of the Fermi level.  
\begin{table*}[!hbt]
\centering
\setlength{\tabcolsep}{4pt}
\renewcommand{\arraystretch}{1.1}
\begin{tabular}{|c|cc|cc|cc|cc|}
\hline 
\multirow{3}{*}{Compound} & \multicolumn{4}{c|}{Lattice parameters (\AA)} & \multicolumn{4}{c|}{Weyl points} \\
\cline{2-9}
 & \multicolumn{2}{c|}{Calculated} & \multicolumn{2}{c|}{Experimental} & \multicolumn{2}{c|}{W1} & \multicolumn{2}{c|}{W2} \\
\cline{2-9}
 & $a$ & $c$ & $a$ & $c$ & ($k_x$, $k_y$, $k_z$) & $E_{W1}$ (meV) & ($k_x$,  $k_y$,  $k_z$) & $E_{W2}$ (meV) \\ \hline
TaAs & 3.4562 & 11.7193 & 3.4368 & 11.644 & (0.0079, 0.5099, 0) & -8.5 & (0.0195, 0.2787, 0.5856) & 2.5 \\
TaP  & 3.3295 & 11.3926 & 3.3184 & 11.374 & (0.0080, 0.5296, 0) & -39.4 & (0.0156, 0.2788, 0.6041) & 26.4 \\
NbAs & 3.4737 & 11.7556 & 3.4500 & 11.670 & (0.0026, 0.4836, 0) & -22.5 & (0.0062, 0.2749, 0.5677) & 7.8 \\
NbP  & 3.3473 & 11.4377 & 3.3340 & 11.420 & (0.0029, 0.5060, 0) & -48.9 & (0.0049, 0.2738, 0.5869) & 22.1 \\ \hline
\end{tabular}
\caption{Comparison of calculated and experimental lattice parameters, together with the positions and energies of Weyl nodes, for TaAs, TaP, NbAs, and NbP. Experimental lattice parameters are from Refs.~\cite{Boller1963,Murray1976,Willerstrom1984,Xu2015}. Weyl point coordinates are given in units of the reciprocal lattice vectors of the conventional unit cell, with the Fermi level set to zero.}
\label{tab:lattice_weyl}
\end{table*}

\begin{figure}[b]
    \centering
    \includegraphics[width=\linewidth]{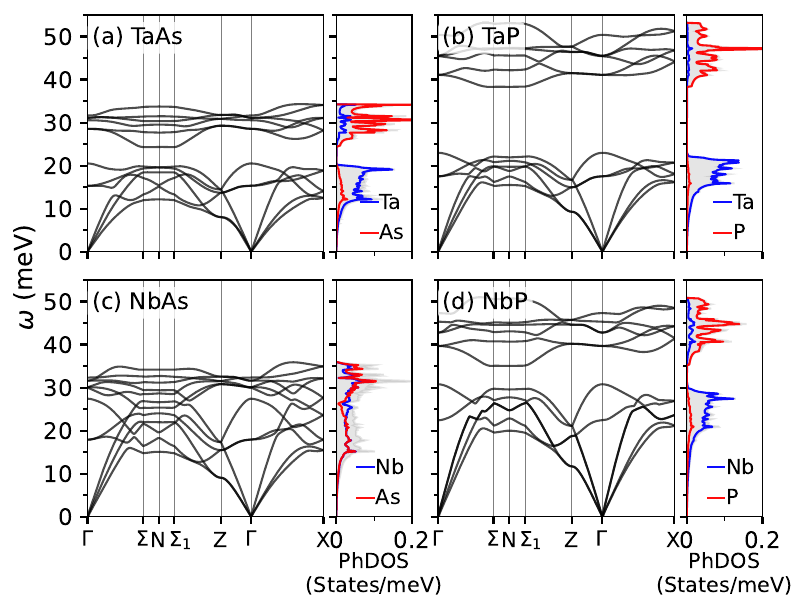}
    \caption{Phonon band structures and phonon density of states (PhDOS) for (a) TaAs, (b) TaP, (c) NbAs, and (d) NbP. Contributions from Ta/Nb and As/P orbitals are shown in blue and red, respectively, while the total PhDOS is represented by the gray shaded area.}
    \label{fig:phonon}
\end{figure}

The magnitude of the band splittings in this family is governed by the strength of SOC. TaAs and TaP show larger splittings due to stronger SOC from the Ta atom, whereas NbAs and NbP exhibit noticeably smaller but still finite splittings attributed to the Nb atom~\cite{Lee2015,Buckeridge2016}. The Fermi-surface (FS) topology reflects these differences (see Fig.~\ref{fig:FS}). TaAs hosts eight banana-shaped hole pockets, while TaP, NbAs, and NbP display additional electron pockets. Among them, NbP exhibits the largest electron and hole pockets, consistent with its enhanced density of states (DOS) at the Fermi level ($N(\ef)$). These features carry important consequences: in Ta-based compounds, strong SOC generally amplifies Berry-curvature-driven responses, including spin Hall conductivity (SHC)~\cite{Sun2016}, and related phenomena~\cite{Zhang2016,Caglieris2018}. Notably, the magnitude of the SHC in these systems is comparable to ordinary 4$d$ and 5$d$ transition metals~\cite{Sun2016,Mishra2025}. By contrast, in Nb-based systems, the larger FS dominates the available phase-space, thereby enhancing the role of e–ph scattering~\cite{Coulter2019}. As shown in Fig.~\ref{fig:FS}(e), the large peak in the nesting function $\zeta_{\bq}$ at $\Gamma$ arises from intra-pocket scattering, while additional peaks are linked to inter-pocket processes. In particular, phonon-mediated inter-pocket transitions along the $\Gamma$-$Z$ and $\Sigma_1$-$Z$ directions are more prominent in NbP and NbAs.
In addition, SOC lifts band degeneracies at certain symmetry TRIM points but preserves the topologically protected crossings between conduction and valence bands at special gapless points, i.e., the Weyl nodes~\cite{Armitage2018}. In all four compounds, the absence of inversion symmetry leads to 12 pairs of Weyl nodes~\cite{Huang2015weyl,Weng2015,Yang2015}. Small variations in lattice parameters, SOC strength, and band structure topology shift the positions and energies of the Weyl points across the TaAs family, as summarized in Table~\ref{tab:lattice_weyl}. These results are consistent with previous theoretical studies, except for TaAs, where Ref.~\cite{Lee2015,Grassano2018} placed both W1 and W2 below $\ef$. This discrepancy likely stems from the sensitivity of $\ef$ determination. Nevertheless, the 11~meV energy offset between the two sets of Weyl nodes in TaAs is in excellent agreement with prior first-principles calculations (14~meV)~\cite{Xu2015,Lee2015,Grassano2018} and angle-resolved photoemission spectroscopy measurements (13~meV)~\cite{Zhang2016}, as elaborated in our recent work~\cite{Liu2025}.  
The phonon dispersions and phonon density of states (PhDOS) are shown in Fig.~\ref{fig:phonon}. Except for NbAs, all compounds display two well-separated frequency regions, with the gap between them increasing as the difference between the transition-metal and pnictogen atomic masses increases. For instance, NbAs shows no gap, while NbP exhibits a small gap of approximately 4~meV. Similarly, the gap widens from about 4~meV in TaAs to nearly 15~meV in TaP. In addition, replacing As with P shifts the optical high-frequency branches upward, with TaAs and NbAs reaching maximum frequencies around 38-42~meV, whereas TaP and NbP extend up to 50-55~meV. The projected PhDOS further reveals that TaAs, TaP, and NbP have low-frequency modes primarily from the transition metal and high-frequency modes from the pnictogen, while in NbAs, the two manifolds overlap, and both species contribute comparably across the entire spectrum. Taken together, the phonon spectra, along with the electronic structure and Fermi-surface geometry, define the phase space relevant for e-ph scattering, and play a central role in shaping charge transport across the TaAs family as discussed next.

\begin{figure}[t]
    \centering
    \includegraphics[width=\linewidth]{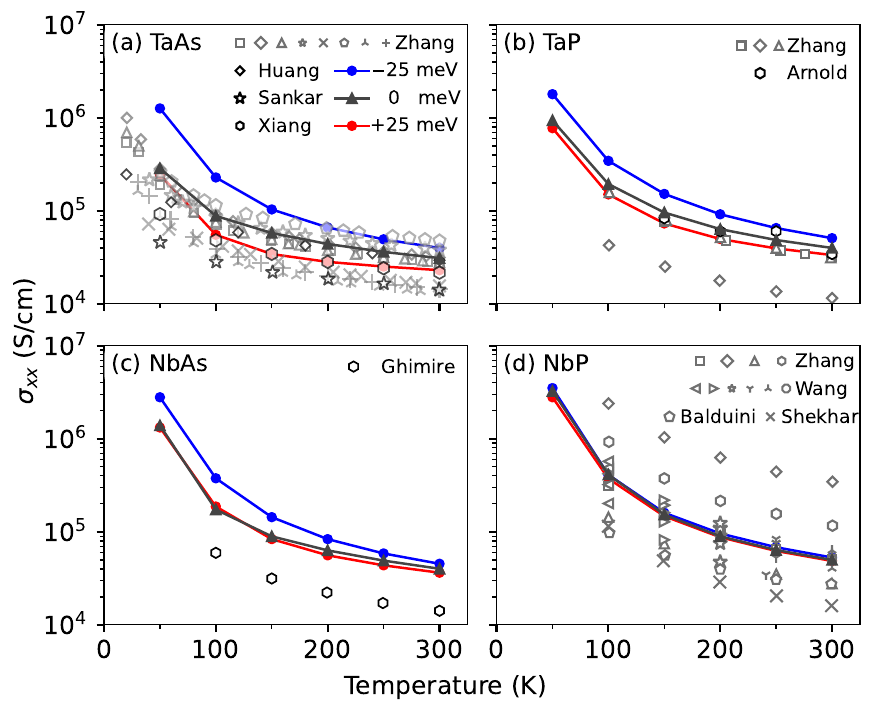}
    \caption{Electrical conductivity $\sigma_{xx}$ as a function of temperature for (a) TaAs, (b) TaP, (c) NbAs, and (d) NbP for three different Fermi level positions: 0 meV (black triangles), $-$25~meV (blue circles), and +25 meV (red circles). Solid symbols denote our calculated results, while hollow symbols represent the experimental data from Refs.~\cite{Zhang2017, Huang2015, Xiang2017, Zhang2015, Arnold2016, Ghimire2015,Shekhar2015,Wang2016,Sankar2018,Balduini2024}. In panels (a), (b), and (d), multiple hollow markers represent different samples reported in Ref.~\cite{Zhang2017} for TaAs, and Refs.~\cite{Zhang2015,Wang2016} for TaP and NbP. The distinct hollow symbols illustrate how the sample quality affects the transport properties.}
    \label{fig:CondT}
\end{figure}

Figure~\ref{fig:CondT} compares the calculated electrical conductivity $\sigma_{xx}$ computed with the IBTE for the undoped ($\ef =\ef^0$) and electron and hole doped ($\ef = \ef^0 \pm 25$~meV) systems with experimental measurements for the TaAs, TaP, NbAs, and NbP compounds. The black (0 meV) curves capture the overall $T$-dependence, with a rapid drop from $\sim 10^5-10^6$ at low-$T$ toward a few $\times 10^4$~S/cm near 300~K, as seen in the experimental data sets. For TaAs and TaP, our results closely reproduce the high-quality experimental data~\cite{Huang2015,Zhang2017,Xiang2017,Zhang2015,Arnold2016}, confirming that in clean samples the transport is predominantly limited by phonon scattering. For NbAs, the calculations exceed the measured conductivities by roughly an order of magnitude~\cite{Ghimire2015}, suggesting sample-dependent extrinsic scattering from impurities or lattice defects which shifts $\ef$ away from the calculated intrinsic conditions. In NbP, the black line lies within the experimental spread over the full temperature range~\cite{Wang2016,Shekhar2015}, again indicating phonon-limited transport in high-quality crystals. In this compound, it has been reported that conduction through surface channels can significantly enhance transport~\cite{Khan2025}. Consequently, the shape and size of the samples may strongly affect the conductivity measured in experiments, and could explain the large differences observed among samples in Refs.~\cite{Wang2016} and \cite{Zhang2015}.

\begin{figure}[!t]
    \centering
    \includegraphics[width=\linewidth]{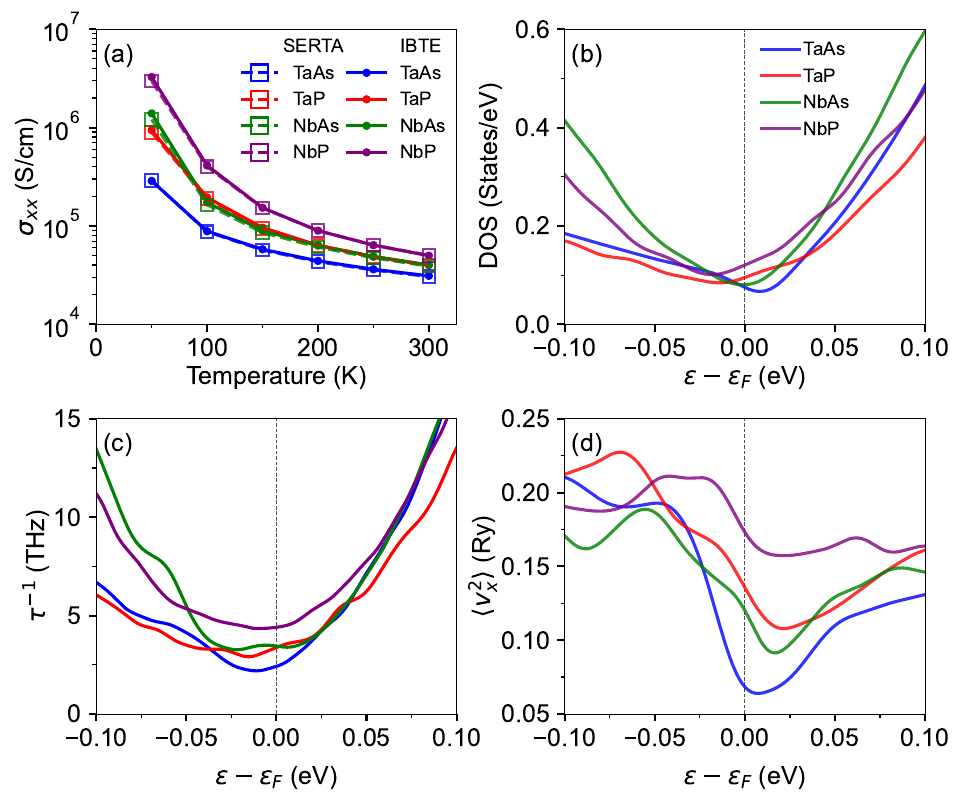}
    \caption{(a) Electrical conductivity $\sigma_{xx}$  from SERTA (dashed lines, open squares) and IBTE (solid lines, filled circles) for TaAs, TaP, NbAs, and NbP in the undoped case. (b) Density of states (DOS), (c) energy-averaged scattering rates at 300~K, and (d) average squared Fermi velocity $\langle v_x^2\rangle$ for all four compounds.}
    \label{fig:cond-analysis}
\end{figure}

Comparing across the series, the calculations reproduce the experimental ordering, NbP has the largest $\sigma_{xx}$ and TaAs the smallest, with TaP and NbAs grouped in the middle (see Fig.~\ref{fig:cond-analysis}(a)). We additionally solved the BTE within self-energy relaxation time approximation (SERTA)~\cite{Liu2025}, and find that the resulting conductivities are indistinguishable from the IBTE values for all four compounds. Once phonon scattering turns on with $T$, all materials are driven toward the same $T$-linear resistive regime, explaining the high-$T$ convergence toward a few $\times 10^4$~S/cm by 300~K. The resistivity curves shown in Fig.~S1~\cite{SI} reproduce both the magnitude and the temperature-dependent slope reported experimentally for TaAs~\cite{Huang2015,Zhang2017,Xiang2017}, TaP~\cite{Arnold2016,Zhang2015}, and NbP~\cite{Zhang2015,Wang2016}, thereby providing further validation of our results.

To further clarify the conductivity trends across the TaAs family, we consider the simplified Boltzmann expression $\sigma_{xx} \propto N(\ef) \langle v^2_{x}\rangle \tau(\ef)$ \cite{Liu2025}, where $N(\ef)$ denotes the DOS at $\ef$, $\langle v^2_{x}\rangle$ is the average squared Fermi velocity along the $x$-direction, and $\tau(\ef)$ represents the energy-averaged relaxation time at $\ef$. The decomposition of $\sigma_{xx}$ into DOS, carrier velocities, and scattering rates for energies within a window around the Fermi level is presented in Fig.~\ref{fig:cond-analysis}(b)-(d) and Table~S1~\cite{SI}. Panel (b) shows that NbP exhibits the largest DOS at $\ef$, which would normally enhance e-ph scattering and suppress conductivity. Panel (c) displays the energy-averaged single-particle quantum scattering rates $\tau^{-1}(\varepsilon)$, which account for all allowed scattering processes regardless of the scattering angle. At $\ef$, this quantity is indeed largest in NbP and smallest in TaAs, with TaP and NbAs lying in between. However, panel (d) reveals that the average squared Fermi velocity $\langle v^2_{x}\rangle$ is also largest in NbP. This is consistent with the weaker SOC of Nb, which reduces band splitting and favors extended linear dispersions. The enhanced velocities offset the combined effects of the DOS and scattering, and are key to sustaining higher conductivities. The near overlap of IBTE and SERTA results in panel (a), indicates that vertex scattering-in corrections, and hence explicit backscattering weighting, are negligible, in line with recent studies on TaAs~\cite{Liu2025,Allemand2025}. The finding that the quantum and transport scattering rates and their Fermi-surface averages are similar for both TaAs and NbP further supports this picture (see Fig.~S2~\cite{SI}). NbP’s larger Fermi velocities compensate its higher quantum and transport scattering rates, thereby sustaining higher conductivity. By contrast, TaAs exhibits stronger Fermi-surface anisotropy and less balanced carrier velocities, leading to lower conductivity even though scattering rates are weaker. Consequently, the ordering of $\sigma_{xx}$ in panel (a) is governed primarily by the balance between the velocity and scattering rate contributions, with phase-space effects playing a secondary role. 

\begin{figure}[!t]
    \centering
    \includegraphics[width=\linewidth]{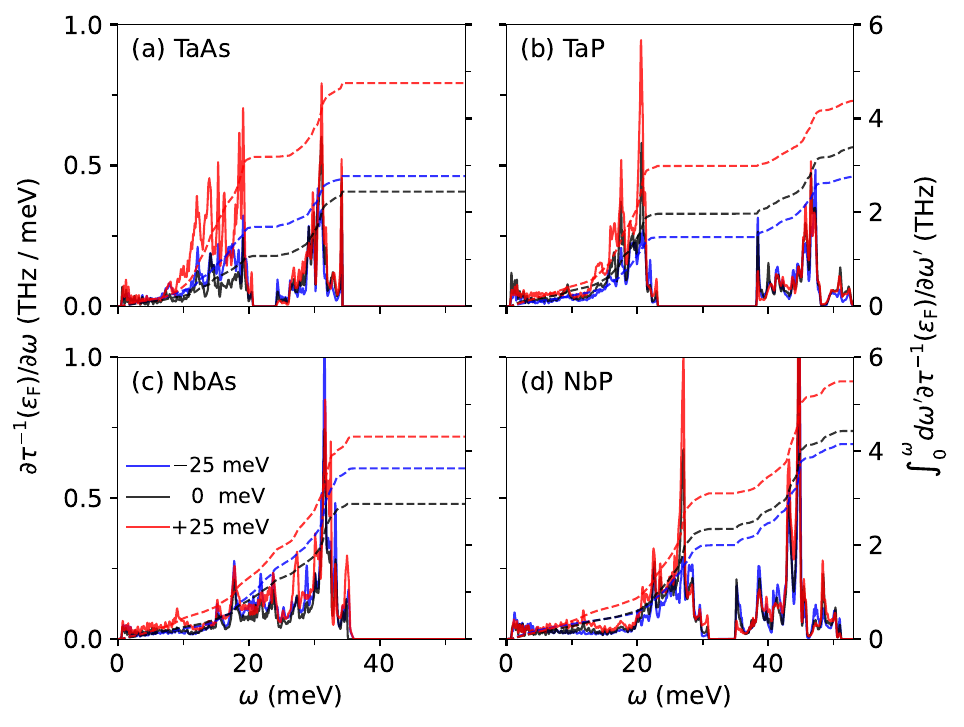}
    \caption{Frequency-resolved scattering rate $\partial \tau^{-1}(\ve)/\partial \omega$ (solid curves, left axis) and cumulative integral $\int_0^{\omega} \rm{d}\omega'\partial \tau^{-1}(\ve)/\partial \omega'$ (dashed curves, right axis) evaluated at the Fermi level and 300~K for (a) TaAs, (b) TaP, (c) NbAs, and (d) NbP.} 
    \label{fig:mode-tau}
\end{figure}

Figure~\ref{fig:mode-tau} presents a frequency-resolved decomposition of the e-ph scattering and its cumulative contribution at the Fermi level. The spectra $\partial\tau^{-1}(\ef)/\partial\omega$ show all four compounds couple to phonons across the full frequency range, with spectral weight coming from both the low- and high-frequency regions. The cumulative integrals make this comparison quantitative. TaAs builds up most slowly and saturates at the smallest value, while NbP accumulates most strongly and ends with the largest integrated scattering. TaP and NbAs lie in between with comparable totals but different distributions (TaP weight concentrated around $15$-$25$ meV, NbAs more broadly spread).

Finally, Figs.~\ref{fig:CondT} and S3~\cite{SI} compare the doping dependence of $\sigma_{xx}$ and scattering rates across the family. TaAs shows a clear electron-hole asymmetry; the undoped case lies between the doped cases, with $\sigma_{xx}$ enhanced under hole doping ($-25$~meV) and suppressed under electron doping ($+25$~meV). Here, the Fermi level sits in a region where small energy shifts strongly modify both the carrier populations (by changing pocket sizes) and the carrier velocities, with the nearby W1 and W2 nodes (see Table~\ref{tab:lattice_weyl}) further amplifying this sensitivity. TaP displays a similar trend, though shifted by the different positions of W1 and W2 nodes. Hole doping increases $\sigma_{xx}$ by enhancing velocity and reducing scattering, while electron doping lowers it as $\ef$ approaches W2, where the scattering rate rises steeply. NbAs exhibits a milder asymmetry; hole doping places $\ef$ near W1 and enhances conductivity through a velocity gain, whereas electron doping leaves $\sigma_{xx}$ essentially unchanged, as the larger DOS is nearly canceled by reduced velocity and increased scattering. In all three cases, the simplified BTE expression captures the doping dependence well, with the conductivity trends tracking the competition between DOS, velocity, and scattering rate (Table~S1~\cite{SI}). In NbP, however, $\sigma_{xx}$ is almost insensitive to $\pm 25$~meV shifts, with the three curves nearly overlapping. Despite W1 and W2 lying relatively close to $\ef$, the large, nearly compensated electron and hole pockets ensure that small Fermi-level shifts preserve carrier balance and do not activate new scattering channels. In this case, the simple BTE picture is insufficient. The robustness of $\sigma_{xx}$ reflects phase-space restrictions imposed by compensation and Fermi-surface geometry, rather than a straightforward competition of DOS, velocity, and scattering.

\section{\label{sec:summary}Summary}

We investigated phonon-limited electrical conductivity in TaAs, TaP, NbAs, and NbP using the \textit{ab initio} Boltzmann transport formalism. Our calculations reveal that despite sharing similar electronic structures, subtle differences in Fermi-surface carrier density, density of states at the Fermi level, and spin-orbit-induced band splittings give rise to distinct transport behaviors. TaAs, with reduced hole pockets, exhibits the weakest scattering but also the smallest carrier velocities, resulting in lower conductivity. By contrast, NbP, with large hole and electron pockets, achieves the highest conductivity, as its higher carrier velocities 
offset its stronger scattering rates. 
Furthermore, the conductivity of TaAs is highly sensitive to carrier doping due to its lower carrier density near the Fermi level, while NbP remains comparatively insensitive. NbAs and TaP lie in between, reflecting their intermediate scattering and velocity characteristics. These results provide a microscopic understanding of phonon-limited transport in the TaAs family and establish a framework for analyzing other Weyl semimetals.

\begin{acknowledgements}
We thank C.-L.~Zhang for providing the experimental raw resistivity data of Ref.~\cite{Zhang2017} and for helpful discussions. 
This work was primarily supported by the Computational Materials Sciences Program funded by the U.S. Department of Energy, Office of Science, Basic Energy Sciences, under Award No. DE-SC0020129. The authors acknowledge the computational resources provided by the Frontera and Stampede3 supercomputers at the Texas Advanced Computing Center (TACC) at The University of Texas at Austin (http://www.tacc.utexas.edu), supported through the Leadership Resource Allocation (LRAC) DMR22004 and the ACCESS allocation TG-DMR180071, and the National Energy Research Scientific Computing Center (a DOE Office of Science User Facility supported under Contract No. DE-AC02-05CH11231).
\end{acknowledgements}

\section*{Author Contributions}
S.M. and Z.L. contributed equally. S.M. wrote the original draft and prepared the figures, while Z.L. carried out the transport calculations and prepared original figures. R.M. conceived and supervised the project, and reviewed and edited the manuscript. All authors participated in the formal analysis and the revision of the paper.


%

\end{document}


\title{Supporting Information: \\Comparative study of phonon-limited carrier transport in the Weyl semimetal TaAs family}

\author{Shashi B. Mishra}
\affiliation{Department of Physics, Applied Physics and Astronomy, Binghamton University-SUNY, Binghamton, New York 13902, USA}
\thanks{These authors contributed equally.}
\author{Zhe Liu}
\affiliation{Department of Physics, Applied Physics and Astronomy, Binghamton University-SUNY, Binghamton, New York 13902, USA}
\thanks{These authors contributed equally.}
\author{Sabyasachi Tiwari}
\affiliation{Oden Institute for Computational Engineering and Sciences, The University of Texas at Austin, Austin, Texas 78712, USA}
\affiliation{Department of Physics, The University of Texas at Austin, Austin, Texas 78712, USA}
\author{Feliciano Giustino}
\affiliation{Oden Institute for Computational Engineering and Sciences, The University of Texas at Austin, Austin, Texas 78712, USA}
\affiliation{Department of Physics, The University of Texas at Austin, Austin, Texas 78712, USA}
\author{Elena R. Margine}
\email{rmargine@binghamton.edu}
\affiliation{Department of Physics, Applied Physics and Astronomy, Binghamton University-SUNY, Binghamton, New York 13902, USA}


\maketitle

\begin{figure*}[!hbt]
    \centering
    \includegraphics[width=0.9\linewidth]{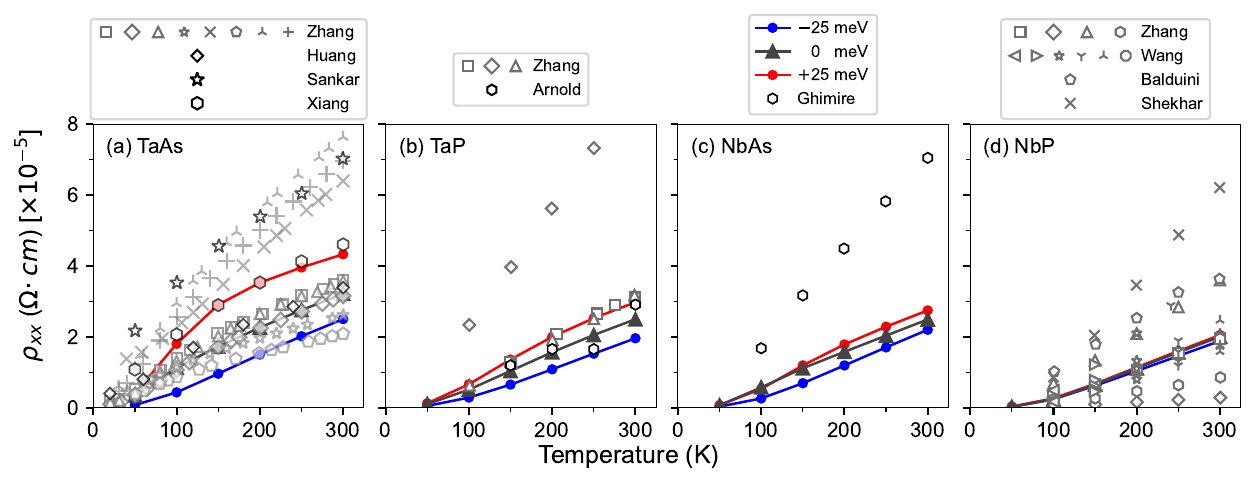}
    \caption{ 
    Electrical resistivity $\rho_{xx}$ as a function of temperature for (a) TaAs, (b) TaP, (c) NbAs, and (d) NbP for three different Fermi level positions: 0 meV (black triangles), $-$25~meV (blue circles), and +25 meV (red circles). Solid symbols denote our calculated results, while hollow symbols represent the experimental data from Refs.~\cite{Zhang2017, Huang2015, Xiang2017, Zhang2015, Arnold2016, Ghimire2015,Shekhar2015,Wang2016,Sankar2018,Balduini2024}. In panels (a), (b), and (d), multiple hollow markers represent different samples reported in Ref.~\cite{Zhang2017} for TaAs, and Refs.~\cite{Zhang2015,Wang2016} for TaP and NbP. The distinct hollow symbols illustrate how the sample quality affects the transport properties.
    }
    \label{fig:resT}
\end{figure*}

\begin{figure*}[!hbt]
    \centering
    \includegraphics[width=0.6\linewidth]{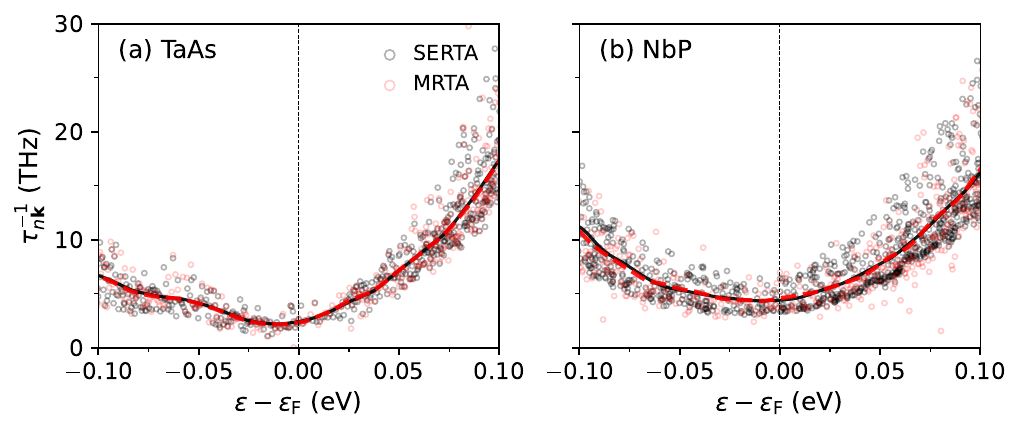}
    \caption{Self-energy relaxation time approximation (SERTA, black) and momentum relaxation time approximation (MRTA, red) scattering rates for (a) TaAs and (b) NbP at 300~K. The energy-averaged scattering rates are shown as black solid line for SERTA and red dashed line for MRTA. }
    \label{fig:SR}
\end{figure*}

\begin{figure*}[!hbt]
    \centering
    \includegraphics[width=0.6\linewidth]{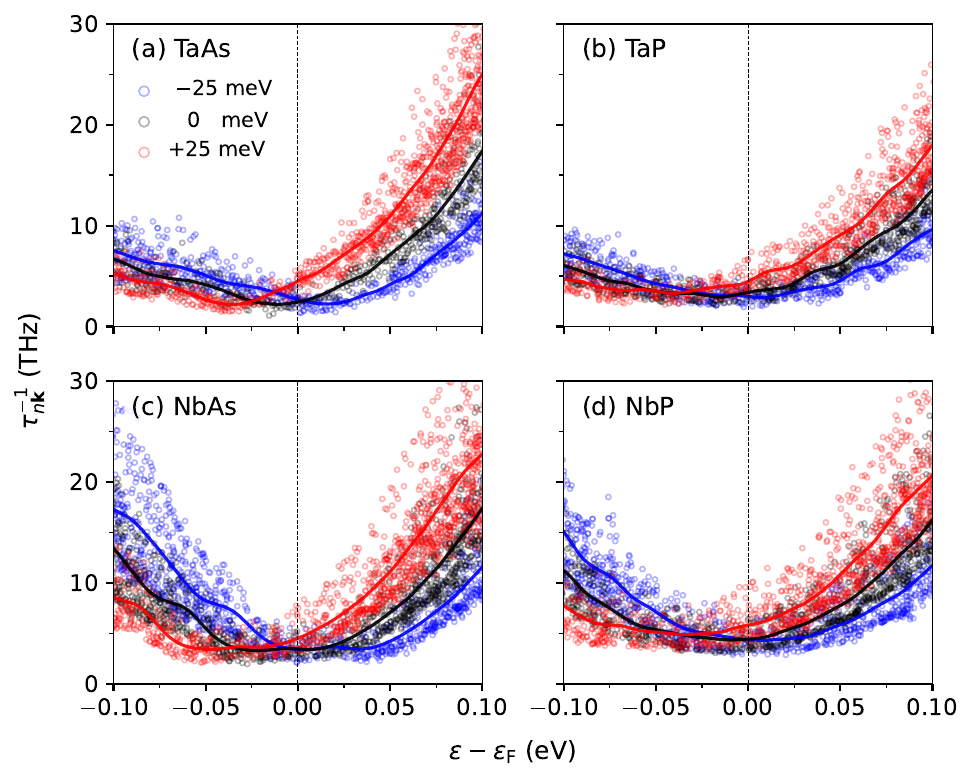}
    \caption{Self-energy relaxation time approximation (SERTA) scattering rates for (a) TaAs, (b) TaP, (c) NbAs, (d) NbP for three different Fermi level positions: 0 meV (black), $-$25~meV (blue), and +25 meV (red) at 300~K. The energy-averaged scattering rates are shown as solid lines. 
    }
    \label{fig:SR}
\end{figure*}

\begin{table}[h!]
\centering
\setlength{\tabcolsep}{8pt}
\renewcommand{\arraystretch}{1.4}
\begin{tabular}{lcccccc}
\hline
\multirow{2}{*}{Material} & $\ef$  
& $N(\ef)$  
& $\langle v_x^2 \rangle$ 
& $\tau^{-1}(\ef)$  
& $\sigma_{xx}^{\text{SIMPLE}}$  
& $\sigma_{xx}^{\text{IBTE}}$ \\
 & (meV) 
& (States/eV/Cell) 
& (Ry) 
& (THz) 
& ($\times 10^4$ S/cm) 
& ($\times 10^4$ S/cm) \\
\hline 
\multirow{3}{*}{TaAs} 
 & $-25$ & 0.1114 & 0.0472 & 2.649 & 7.0082 & 3.9936 \\
 & 0     & 0.0746 & 0.0108 & 2.417 & 1.8602 & 3.1124 \\
 & $+25$ & 0.1041 & 0.0114 & 4.519 & 1.3848 & 2.3086 \\
\hline
\multirow{3}{*}{TaP} 
 & $-25$ & 0.0897 & 0.1714 & 2.979 & 14.146 & 5.0977 \\
 & 0     & 0.0951 & 0.1350 & 3.371 & 10.416 & 4.0010 \\
 & $+25$ & 0.1250 & 0.1084 & 4.437 & 8.3366 & 3.3570 \\
\hline
\multirow{3}{*}{NbAs} 
 & $-25$ & 0.1160 & 0.1512 & 3.365 & 14.213 & 4.5398 \\
 & 0     & 0.0806 & 0.1197 & 3.452 & 7.6683 & 4.0118 \\
 & $+25$ & 0.1400 & 0.0972 & 4.491 & 8.3088 & 3.6393 \\
\hline
\multirow{3}{*}{NbP} 
 & $-25$ & 0.1050 & 0.2101 & 4.358 & 13.801 & 5.2983 \\
 & 0     & 0.1200 & 0.1728 & 4.402 & 12.955 & 4.9896 \\
 & $+25$ & 0.1700 & 0.1574 & 5.788 & 12.632 & 4.8708 \\
\hline 
\end{tabular}
\caption{Electrical conductivities for TaAs, TaP, NbAs, and NbP calculated with the simplified BTE model and the IBTE approach for different Fermi-level shifts at 300~K.}
\end{table}

\newpage


%